\documentclass[aps,prl,showpacs,superscriptaddress,prbib, twocolumn]{revtex4}
\usepackage{graphicx}
\usepackage{float}
\newcommand{\Fig}[1]{Fig.~\ref{#1}}

\begin{document}

\author{Clifford W. Hicks}
\affiliation{Scottish Universities Physics Alliance (SUPA), School of Physics and Astronomy, University of St.
Andrews, St. Andrews KY16 9SS, United Kingdom}
\author{Alexandra S. Gibbs}
\affiliation{Scottish Universities Physics Alliance (SUPA), School of Physics and Astronomy, University of St.
Andrews, St. Andrews KY16 9SS, United Kingdom}
\affiliation{School of Chemistry and EaStCHEM, University of St. Andrews, North Haugh, St. Andrews KY16 9ST,
United Kingdom}
\author{Andrew P. Mackenzie}
\email{apm9@st-andrews.ac.uk}
\affiliation{Scottish Universities Physics Alliance (SUPA), School of Physics and Astronomy, University of St.
Andrews, St. Andrews KY16 9SS, United Kingdom}
\author{Hiroshi Takatsu}
\affiliation{Department of Physics, Tokyo Metropolitan University, Tokyo 192-0397, Japan}
\author{Yoshiteru Maeno}
\affiliation{Department of Physics, Graduate School of Science, Kyoto University, Kyoto 606-8502, Japan}
\author{Edward A. Yelland}
\affiliation{Scottish Universities Physics Alliance (SUPA), School of Physics and Astronomy, University of St.
Andrews, St. Andrews KY16 9SS, United Kingdom}
\affiliation{ SUPA, School of Physics and Astronomy, and Centre for Science at Extreme Conditions, University
of Edinburgh, Mayfield Road, Edinburgh EH9 3JZ, United Kingdom }

\title{Quantum oscillations and high carrier mobility in the delafossite PdCoO$_2$}

\date{23 July 2012}

\begin{abstract}

We present de Haas-van Alphen and resistivity data on single crystals of the delafossite PdCoO$_2$. At 295~K
we measure an in-plane resistivity of 2.6~$\mu\Omega$-cm, making PdCoO$_2$ the most conductive oxide known.
The low-temperature in-plane resistivity has an activated rather than the usual $T^5$ temperature dependence,
suggesting a gapping of effective scattering that is consistent with phonon drag. Below 10~K, the transport
mean free path is $\sim$20~$\mu$m, approximately $10^5$ lattice spacings and an astoundingly high value for
flux-grown crystals. We discuss the origin of these properties in light of our data.

\end{abstract}

\maketitle

A number of discoveries, for example graphene and the superconductivity in the sodium cobaltate system, have
focused attention on the properties of two-dimensional conduction on triangular lattices. One of the most
prominent families of such materials is the delafossites ABO$_2$, where the A atoms (Pt, Pd, Ag or Cu) are
directly bonded to each other in triangular-lattice sheets, separated by BO$_2$ layers. Possibilities for B
include Cr, Co, Fe, Al and Ni~\cite{Marquardt05, Shannon71}. The delafossite family includes band insulators and
semiconductors~\cite{Shannon71, Rogers71}, transparent semiconductors~\cite{HosonoGroup}, metallic
magnets~\cite{Takatsu10}, candidate thermoelectrics~\cite{Singh07}, and magnetic and magneto-electric
insulators~\cite{Wang08}. There are also nonmagnetic metals that present a combination of strong anisotropy,
high carrier density and high mobility that is unique for bulk materials~\cite{Rogers71, Carcia80}. In the long term,
the delafossites may be a candidate family for multilayer technologies benefiting from their broad range of
physical properties. High purity is a prerequisite for such developments, and so the extremely high
conductivity of nonmagnetic delafossite metals is worthy of investigation. Here, we concentrate on the most
extensively studied material, PdCoO$_2$.

PdCoO$_2$ crystallizes in the space group $R \bar{3} m (D_{3d}^5)$. The lattice parameters of the hexagonal
unit cell at 298~K are $a=b=2.830$~\AA, and $c=17.743$~\AA~\cite{Takatsu07}. Single crystals have been grown
by several groups~\cite{Tanaka96, Takatsu07, Noh09_ARPES}. The Co atoms have formal valence $+3$ and
configuration $3d^6$, and calculations place the Fermi level $E_F$ between filled pseudo-$t_{2g}$ and empty
pseudo-$e_g$ levels~\cite{Seshadri98, Eyert08, Kim09}. The Pd sheets therefore dominate the conductivity,
leading to high transport anisotropy. The Pd atoms have formal valence $+1$, in  configuration $4d^{9-x}5s^x$.
A key issue in understanding the physics of PdCoO$_2$ is the relative importance of the $4d$ and $5s$ states
to the conduction. We will return to this in light of our data.

PdCoO$_2$ is nonmagnetic, and so shows none of the frustration effects commonly associated with magnetism in
triangular lattices. Band structure calculations~\cite{Eyert08, Ong10} and angle-resolved photoemission
spectroscopy (ARPES) data~\cite{Noh09_ARPES} indicate that it has a broad, half-filled conduction band,
leading to a single, nearly two-dimensional Fermi surface (FS) with a rounded hexagonal cross section. This is
different from graphene or graphite, since the Pd atoms form a simple triangular array rather than a honeycomb
lattice.

At room temperature, PdCoO$_2$ is the most conductive oxide known. To measure the in-plane resistivity
$\rho_{ab}$, we cut three needle-like samples from a 6~$\mu$m-thick single crystal, of length $\sim400$~$\mu$m
and widths 39, 42 and 120~$\mu$m. The measured resistances of all samples were consistent with
$\rho_{ab}$(295~K)$\:=2.6(2)$~$\mu\Omega$-cm: two and four times, respectively, less resistive than other
oxides known for their high conductivity, SrMoO$_3$ and ReO$_3$~\cite{Nagai05, Pearsall74}. Among the
elemental metals, only Cu, Ag and Au have lower resistivities (1.7, 1.6 and 2.2~$\mu\Omega$-cm, respectively).
The high conductivity of PdCoO$_2$ is especially surprising considering that the CoO$_2$ layers do not
contribute itinerant carriers: the carrier density is 30\% that of Cu. 

\begin{figure}
\includegraphics[width=3.25in]{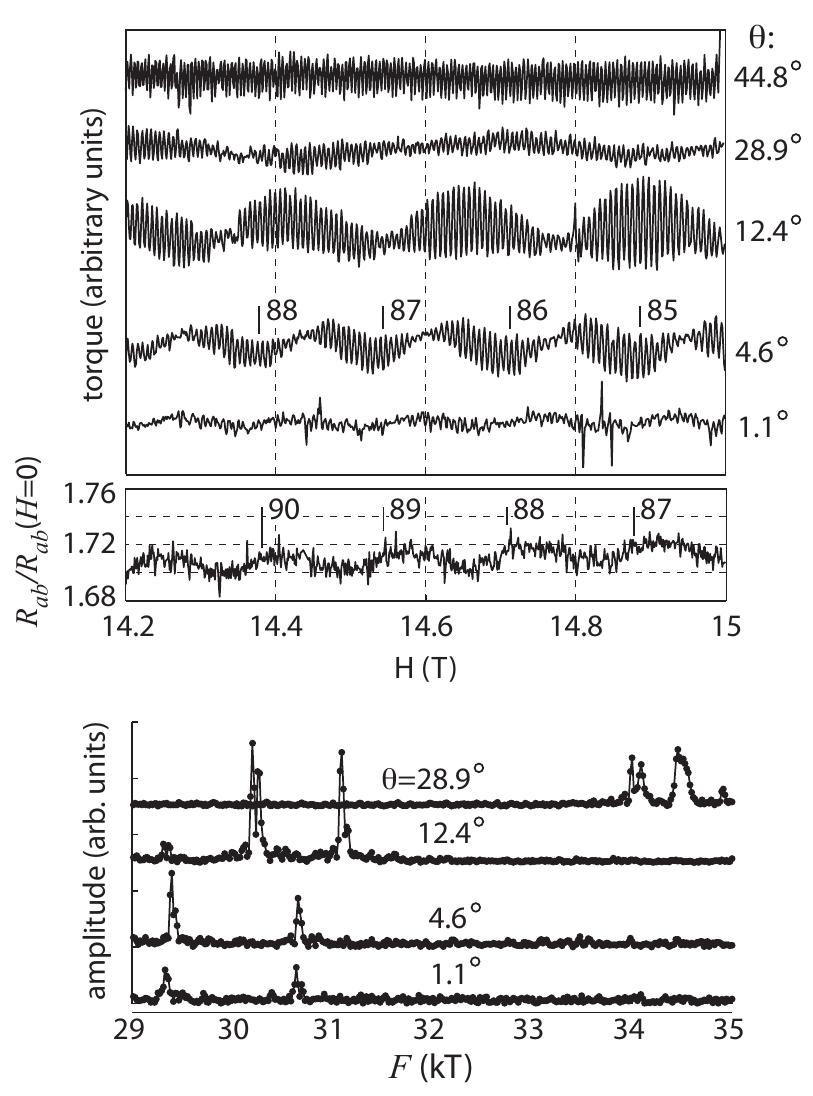}
\caption{\label{sampleTorque} Upper panel: torque curves against $H$, at the indicated angles $\theta$ between
$\mathbf{H}$ and the $c$-axis, and magnetoresistance of a resistivity sample. The ticks on this curve
and the 4.6$^\circ$ torque curve label the cycle number of the difference frequency: the troughs in the
resistance oscillations match up with points where the torque oscillation amplitude (and likely magnetic
interaction) is largest.  Lower panel: Fourier transforms over the field range 8.1-15~T.} 
\end{figure}

\begin{figure*}
\includegraphics[width=6.5in]{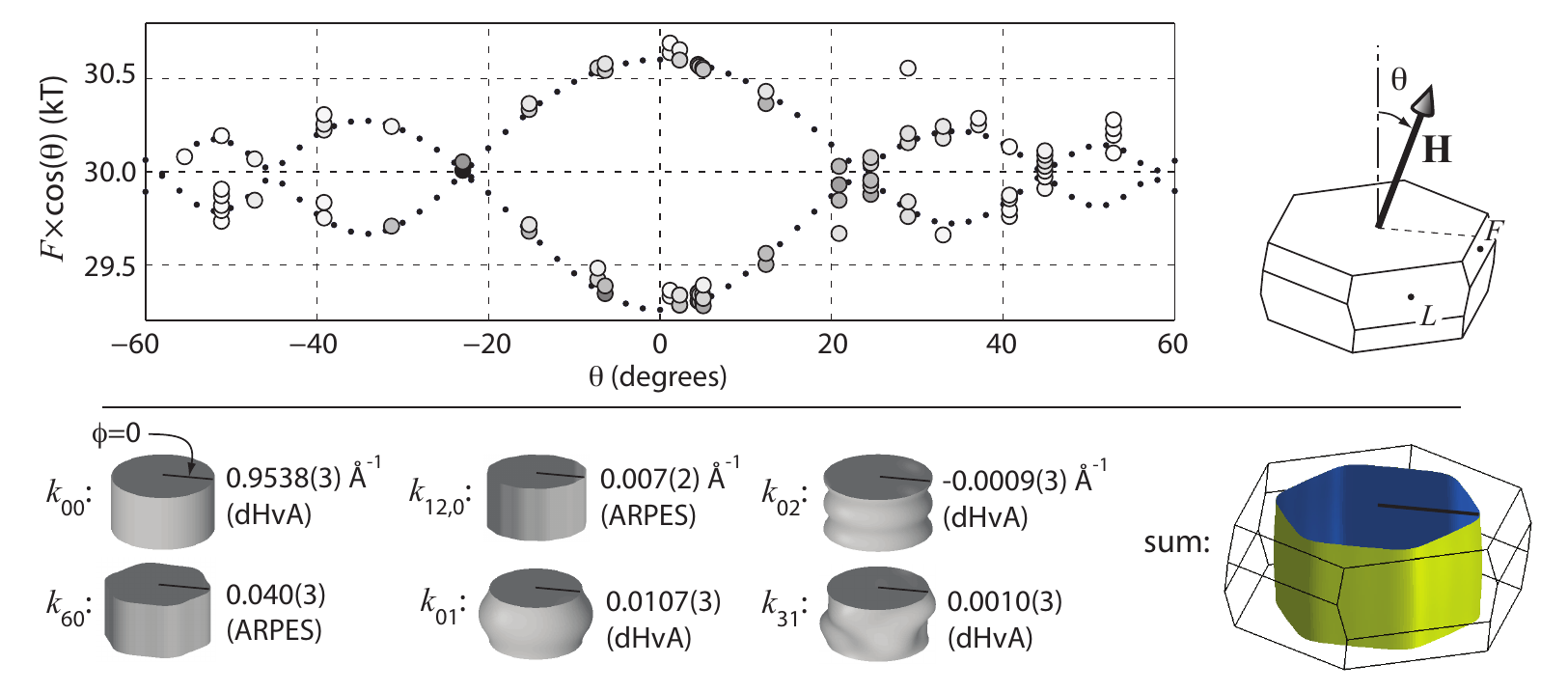}
\caption{\label{freqFit}Top: Observed dHvA frequencies $F$, obtained by Fourier transform over the field range
8.1 to 15~T (circles; darker shading indicates larger amplitude), and a fit (small points). Bottom: the
separate harmonic components included in the fit, and their amplitudes. $k_{60}$ and $k_{12,0}$ are
obtained from ARPES data~\cite{Noh09_ARPES}, and $k_{00}$, $k_{01}$, $k_{02}$ and $k_{31}$ from the fit.
$k_{01}$ gives the overall Yamaji-type form to $F(\theta)$ and $k_{02}$ an asymmetry between the upper and
lower frequencies. $k_{31}$ gives an asymmetry between $\theta>0$ and $\theta<0$, which is allowed by the lack
of mirror symmetry in the crystal structure. The Fermi surface that results from summing these components is
shown at right.}
\end{figure*}

To study the origin of this high conductivity, we combined a de Haas-van Alphen (dHvA) study, using a
piezo-resistive torque method~\cite{Cooper03}, with high-resolution resistivity measurements. The crystals
were grown as described in Ref.~\cite{Takatsu07}.  They are clean: the resistivity samples have
$\rho_{ab}$(295~K)/$\rho_{ab}(0) \approx 250$. Torque measurements were carried out between 50~mK and 20~K, on
a crystal $\approx120$~$\mu$m across and 30~$\mu$m thick; example torque data are shown in \Fig{sampleTorque}.
That the measurement current was not heating the samples was checked with simultaneous measurement of dHvA
oscillations in Sr$_2$RuO$_4$, a renormalized Fermi liquid with well-known quasiparticle masses. Using a
rotatable sample stage, the field $\mathbf{H}$ could be rotated by an angle $\theta$ from the crystal $c$-axis
towards the $F$ symmetry point of the first Brillouin zone (as illustrated in \Fig{freqFit}). The temperature
dependence of the dHvA effect was studied at $\theta=-6^\circ$, and the $\theta$ dependence at $T=0.7$~K. 

The data in \Fig{sampleTorque} are consistent with PdCoO$_2$ being a high purity, high carrier concentration,
quasi-2D metal. At each angle $\theta$ we observe two large dHvA frequencies that differ by only $\sim$3\%,
and that vary approximately as $1/\cos(\theta)$, confirming the overall cylindrical form of the FS. We also
observe significant magnetic interaction (MI), an effect seen in the highest purity metals~\cite{Shoenberg}.
Most torque curves show a common characteristic of MI, a strong difference frequency amplitude, visible in
\Fig{sampleTorque} as an oscillation matching the beating of the two high frequencies. The difference
frequency also appears in the magnetoresistance, ruling out torque interaction as its origin.

Considerable insight into the physics of PdCoO$_2$ can be obtained from detailed analysis of the dHvA data.
Following~\cite{Bergemann03} and~\cite{Grigoriev10}, the corrugations on nearly-cylindrical Fermi surfaces can
be expanded into harmonic components with amplitudes $k_{\mu\nu}$:
\[
k_F(\phi,k_z) = \sum_{\mu,\nu}k_{\mu \nu} \cos{\mu \phi} \left\{ \begin{array}
	{l}\cos \nu d k_z \\
	\sin \nu d k_z \end{array} \right.,
\]
where $k_z$ is the $z$ component of the wavevector, $d=c/3$ the interlayer spacing, and $\phi$ the azimuthal
angle. $k_{00}$ is the average radius of the FS, and the remaining $k_{\mu\nu}$ are the amplitudes of 
harmonic components of the corrugation. The components considered in this paper are illustrated in
\Fig{freqFit}.  The triangular lattice of the Pd sheets allows components with $\mu=0,6,12,...$ with an even
({\it i.e.} cosine) dependence on $k_z$. In addition, because PdCoO$_2$ lacks mirror symmetry along the $c$-axis, 
$\mu=3,9,15,...$ are allowed with odd ({\it i.e.} sine) $k_z$ dependence. 

Since $k_z$-independent warpings cannot be obtained from dHvA frequencies, we set $k_{60}=0.040(3)$ and
$k_{12,0}=0.007(2)$~\AA$^{-1}$ to match the ARPES Fermi surface~\cite{Noh09_ARPES}. With these
amplitudes fixed, we then allow $k_{00}$, $k_{01}$, $k_{02}$ and $k_{31}$ to vary to obtain the best match to
the dHvA frequencies, setting all higher-order amplitudes to zero. The FS cross-sectional areas in the fit to
the data were calculated by numerical integration~\cite{FSareaFootnote}. The fit and the resulting FS are
shown in \Fig{freqFit}.  The data constrain the relative signs of $k_{01}$, $k_{02}$ and $k_{31}$; $k_{01}$ is
taken to be $>0$ to match electronic structure calculations.

Numerical integration of the volume enclosed by the FS yields a carrier density of 1.005(1) per formula unit,
based on room-temperature values for $a$ and $b$. If $a$ and $b$ shrink by $\sim$0.25\% at low temperatures
then the carrier density becomes almost exactly one: the FS of PdCoO$_2$ arises, to high precision, from
a single half-filled band.

As mentioned above, a key issue for understanding PdCoO$_2$ is the relative importance of the Pd $4d$ and
$5s$ orbitals in states near the Fermi level. The dHvA masses [$m^* \: \equiv (\hbar^2/2\pi) dA/dE$, where $A$
is the FS cross-sectional area] provide the first piece of evidence that the $5s$ component is more important
than has usually been assumed. Our data yield $m^*=1.45(5)m_e$ and $1.53(3)m_e$ for the 29.3 and
30.7~kT orbits, respectively (at $\theta = -6^\circ$)~\cite{Supp}, corresponding to an electronic specific
heat  of
approximately 1~mJ/mol-K$^2$, in fairly good agreement with a measured value of
1.28~mJ/mol-K$^2$~\cite{Takatsu07, Supp}. These masses would be very low for a
$d$-orbital-derived FS. For comparison, Pd metal (which has atomic spacing 2.75~\AA, just 3\% less than in
PdCoO$_2$) has a mainly $5s$-derived electron-like Fermi pocket, compensated by a $4d$-derived open hole
surface, and the dHvA masses of these surfaces are $\sim2.0$ and 5--10$m_e$, respectively~\cite{Mueller70,
Dye81}. The PdCoO$_2$ dHvA masses are much closer to that of the Pd $5s$ than the Pd $4d$ surface. 

The form of the FS warping also emphasizes the role played by the $5s$ component. The dominant
warping is $k_{01}$, suggesting interlayer transport by direct Pd-Pd overlap, which would rely on the extended
$5s$ component to cross the 6~\AA \ interlayer separation. Hopping via the Co atoms would allow transport
through the more compact $d$ orbitals, but connects interlayer next-nearest-neighbor Pd sites (through the
path Pd-O-Co-O-Pd, illustrated in Fig.~5b of~\cite{Ong10}), leading to a large $k_{31}$. Published
calculations indeed show bumps on the FS with significant Co character that give a large $k_{31}$~\cite{Ong10,
Eyert08}. The dHvA data, however, show that $k_{01}$ is far larger than $k_{31}$.

Hybridization between the Pd $4d_{z^2}$ and $5s$ orbitals has been discussed in~\cite{Tanaka98, Seshadri98,
Rogers71}. Reports on electronic structure calculations have tended to mention the likelihood of a $5s$
component at $E_F$ but emphasize the $4d$ contribution~\cite{Eyert08, Seshadri98, Ong10, Kim09}. However in a
calculation of PdCoO$_2$ without the CoO$_2$ spacer layers, {\it i.e.} bare Pd sheets, a $5s$ manifold that
straddles $E_F$ is clearly visible (Fig.~9 of Ref.~\cite{Seshadri98}), and remains apparent in full PdCoO$_2$
band structures. 

To analyze our data and compare with previous calculation, we performed electronic structure calculations on
PdCoO$_2$ using the general potential linearized augmented plane wave method as implemented in the WIEN2K
package. We calculated 20,000 $k$ points in the full Brillouin zone, and studied the effects of an on-site
repulsion $U_{\mathrm{eff}}$ at the Co atoms. For $U_{\mathrm{eff}}=0$, we recover previously reported
results, and find dHvA masses of around $1.9m_e$, larger than the measured value. Introducing a modest
$U_{\mathrm{eff}}$ reduces the calculated masses towards the experimental values and simultaneously gives a
ratio between $k_{31}$ and $k_{01}$ that is in better agreement with experiment (Table~1). The high
sensitivity of the in-plane masses, which arise from Pd-Pd bonding, to the Co site Coulomb repulsion appears
surprising. A possible interpretation is that the Pd $5s$ component is extended enough to be sensitive to the
charge distribtuion in the CoO$_2$ layers; this should be the subject of further investigation.
\begin{table}[b]
\caption{LDA+U results, and experimental data. $U_{\mathrm{eff}}$ is the effective on-site Coulomb repulsion
at the Co sites. $m^*_\Gamma$ is the dHvA mass for the orbit in the $k_z=0$ plane, and $m^*_Z$ the
$k_z=\pm\pi$ plane.}
\begin{tabular*}{\columnwidth}{@{\extracolsep{\fill}}llllll}
\hline\hline
 & $m^*_\Gamma$ & $m^*_Z$ & $k_{01}$ (\AA$^{-1}$) & $k_{31}$ (\AA$^{-1}$) & $k_{31}/k_{01}$\\
\hline
$U_{\mathrm{eff}}=0$~eV & $1.90m_e$ & $1.82m_e$ & 0.0039 & 0.0172 & 4.4 \\
$U_{\mathrm{eff}}=5$~eV & 1.66 & 1.52 & 0.0181 & 0.0029 & 0.2 \\
experiment & 1.53 & 1.45 & 0.0107(3) & 0.0010(3) & 0.1 \\
\end{tabular*}
\end{table}

\begin{figure}
\includegraphics[width=3.25in]{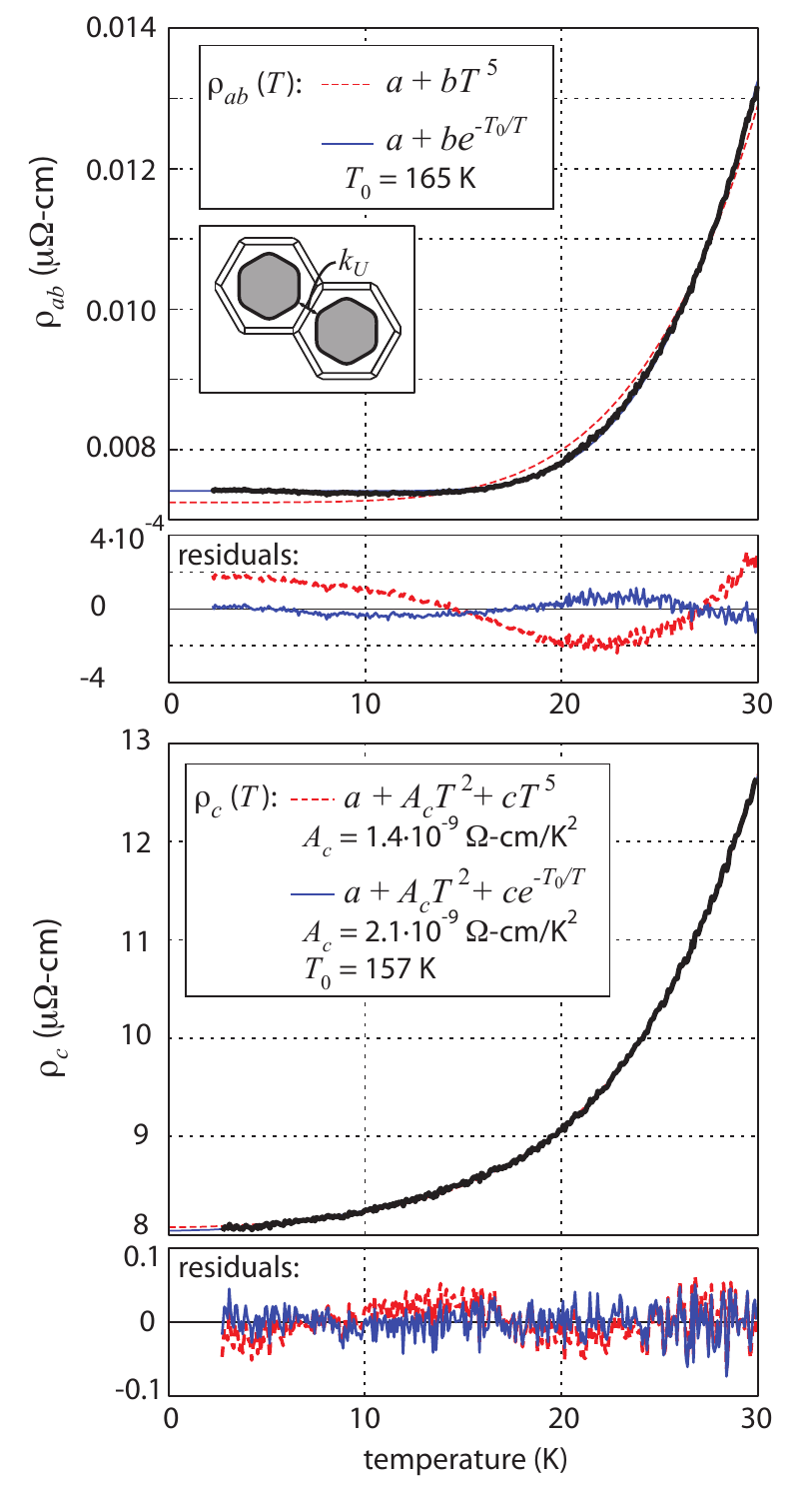}
\caption{\label{resistivity} In-plane PdCoO$_2$ resistivity $\rho_{ab}(T)$ (top), and $c$-axis resistivity
$\rho_c(T)$ (bottom), with fits. At low $T$, $\rho_c$ has a $T^2$ dependence, while $\rho_{ab}$ is nearly
temperature-independent. The inset in the $\rho_{ab}$ panel illustrates $k_U$, the minimum wavenumber for
Umklapp scattering.}
\end{figure}

Although conduction by $5s$ states might account for the scale of the conductivity of PdCoO$_2$ at room
temperature, the low temperature resistivity has additional intriguing features.
Measurement is challenging because of the tiny resistance of typical samples. By using transformers mounted
at 1~K to provide low-noise (150~pV/$\sqrt{\mathrm{Hz}}$) passive amplification, we obtained the data shown in
\Fig{resistivity}. The difference between $\rho_c$ and $\rho_{ab}$ is striking: while $\rho_c$ shows a $T^2$
dependence at low temperatures, $\rho_{ab}$ becomes almost temperature-independent.

Below 30~K, $\rho_{ab}(T)$ is much better described by an activated form than the usual electron-phonon $T^5$
form (\Fig{resistivity}). This may be a consequence of phonon drag, in which the phonon population is pushed
out of equilibrium by charge carrier flow. It is a well-known phenomenon in low-temperature thermopower, but
has been seen in resistivity only in the alkali metals~\cite{Bass90}. In clean materials momentum is
transferred to the lattice through Umklapp processes. For the alkali metals and for in-plane conduction in
PdCoO$_2$, the Fermi surface is closed, resulting in an activation energy $k_BT_U=\hbar c k_U$, where $k_U$ is
the minimum wavenumber for Umklapp processes (illustrated in \Fig{resistivity}) and $c$ is the speed of sound. 
For $T \ll T_U$, electron-phonon scattering transfers momentum between electrons and phonons but does not
relax the momentum of the combined system. $k_U \approx 0.5$~\AA$^{-1}$ in PdCoO$_2$. Fits to $\rho_{ab}(T)$
for two samples yield activation temperatures of 165 (\Fig{resistivity}) and 168~K (not shown), giving $c
\approx 4400$~m/s. This is in fair agreement with specific heat measurements, where the magnitude of the $T^3$
term, 0.0619~mJ/mol-K$^4$~\cite{Takatsu07}, yields, for isotropic sound propagation, $c=3650$~m/s. This
analysis works equally well for potassium~\cite{Kfoot}.

Fermi liquids usually have a $T^2$ low-temperature resistivity due to electron-electron scattering.  $A_c$ in
$\rho_c(T) = \rho_c(0) + A_c T^2$ is $2 \cdot 10^{-9}$~$\Omega$-cm/K$^2$. $\rho_{ab}(T)$ shows a very small
resistivity upturn at low $T$, and the upper limit we can place on $A_{ab}$ is $1.0 \cdot
10^{-12}$~$\Omega$-cm/K$^2$. This is not a particularly small value for metals with electronic specific
heats comparable to PdCoO$_2$~\cite{AlResistivity}, but it is significantly smaller than $A_c$ scaled by the
low-$T$ resistive anisotropy ($\rho_c(0)/\rho_{ab}(0) = 1000$). In the alkali metals K and Na the
closed FS geometry is known to partially suppress electron-electron Umklapp processes, reducing $A$ by about a
factor of ten~\cite{Kaveh84}. Such a suppression is also likely to apply to $A_{ab}$ of PdCoO$_2$.

The combination of resistivity and dHvA data also shows that transport electron-phonon coupling in PdCoO$_2$
is very small over an extended temperature range. In Ref.~\cite{Takatsu07}, $\rho_{ab}(T)$ of
PdCoO$_2$ for $0<T<500$~K is fitted to a sum of the Bloch-Gruneisen form and an Einstein mode contribution.
These fits return Debye and Einstein temperatures $\Theta_D$ and $\Theta_E$, and transport electron-phonon
couplings scaled by the the plasma frequency squared, $\lambda_{D, \mathrm{tr}}/\Omega_p^2$ and $\lambda_{E,
\mathrm{tr}}/\Omega_p^2$~\cite{Allen93}. The dHvA data here allow determination of the plasma frequency:
$\Omega_{p, xx}^2=ne^2/m\epsilon_0 = (7.2(1)\cdot10^{15}$ sec$^{-1}$)$^2$. Performing the same fit
over the range $60<T<300$~K, where the lower bound excludes the region strongly affected by phonon drag, we
obtain $\Theta_D=329$~K and $\Theta_E=1106$~K, in good agreement with~\cite{Takatsu07}.
$\lambda_{D,\mathrm{tr}}$ and $\lambda_{E, \mathrm{tr}}$ come to 0.043 and 0.023, respectively, which are
exceptionally low values: Cu has $\lambda_{D,\mathrm{tr}}=0.13$ and Ca, a metal with a particularly low dHvA
mass of $\sim0.6m_e$, 0.05~\cite{Allen87}. These low couplings may also be due to the FS topology, through a
partial suppression of Umklapp processes at temperatures $T \sim T_U$. Careful calculation would be required
to test this hypothesis.

Our measured residual resistivities $\rho_{ab}(T \rightarrow 0)$, between 0.007 and 0.008~$\mu\Omega$-cm for
two samples, are very low, corresponding to a transport mean free path of 20~$\mu$m, or, in a standard
interpretation, a defect density in the Pd layers of one in $10^5$~\cite{Dingle}. This is a remarkable value
for flux-grown crystals. To set it in context, the fractional quantum Hall effect has recently been observed
in ZnO-based heterostructures, where, after years of research, the transport mean free path has been increased
to $\sim$1.5~$\mu$m~\cite{Tsukazaki10}. This is considerably less than what we report here for a naturally
two-dimensional material, motivating the study of delafossites as a natural material for multilayer oxide
technologies.

In summary, we have observed exceptionally high conductivity in PdCoO$_2$ and mapped its Fermi surface in
detail using the de Haas-van Alphen effect. Our data suggest that the conductivity of PdCoO$_2$ is due
primarily to Pd $5s$ overlap. For in-plane transport, the contributions to resistivity from electron-phonon,
electron-electron and electron-impurity scattering are all anomalously low. The effect of the closed Fermi
surface topology on Umklapp processes seems to play a role and is worthy of investigation. It is also
possible, however, that scattering is suppressed by some non-standard mechanism not considered in the analysis
we have presented. Our results provide motivation for further work on this fascinating material.

We acknowledge useful discussions with J.W. Allen, C. Varma, C.A. Hooley, D.J. Singh, and S. Yonezawa, and
practical assistance from P.A. Evans. This work has made use of the resources provided by the Edinburgh Compute
and Data Facility (ECDF), which is partially supported by the eDIKT initiative. We acknowledge funding from
the UK EPSRC, the MEXT KAKENHI (No. 21340100), the Royal Society and the Wolfson Foundation.

\newpage

\section{Supplemental Material}

\begin{figure}[pb]
\includegraphics[width=3.25in]{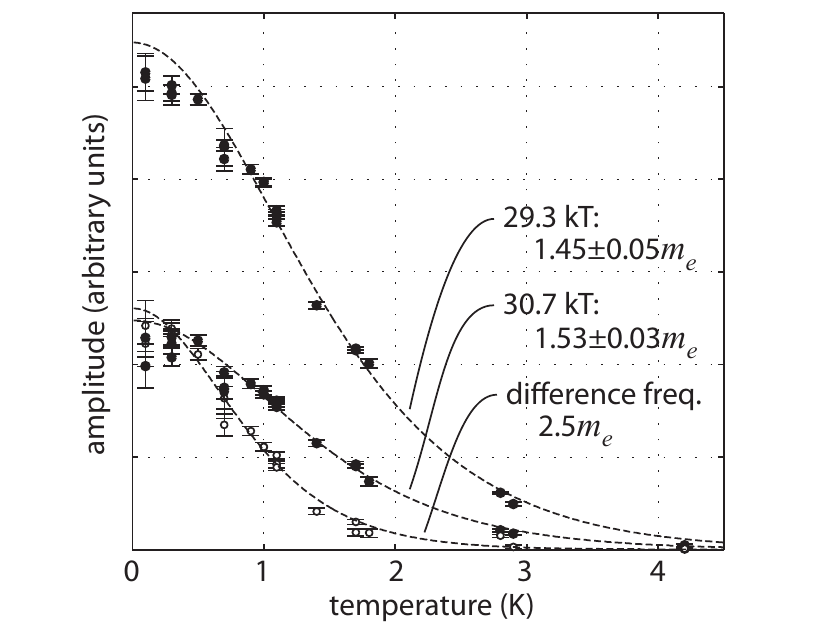}
\caption{\label{LKfits}Fits of dHvA amplitudes at $\theta=-6^\circ$, calculated over the field range
12.86-15~T, to the Lifshitz-Kosevich form, for the indicated frequencies. The masses resulting from the fits
are indicated in the figure.}
\end{figure}

{\it \noindent Mass determination.} The Lifshitz-Kosevich fits used to determine the cyclotron masses at
$\theta = -6^\circ$ are shown in \Fig{LKfits}. The difference frequency probably appears as a result of
magnetic interaction. If data below 1.3~K are excluded from the 29.3 and 30.7~kT fits, out of concern that
magnetic interaction might affect the observed amplitudes, then masses of 1.50 and $1.66m_e$ are obtained,
respectively.
\\

{\it \noindent Specific heat.} The electronic specific heat is given by $\gamma \equiv C_{el}/T = \pi k_B^2 m^*
/ 3 d \hbar^2 \times V$, where $V$ is the volume per mole. If $m^*$ is taken to be $1.50m_e$ at all $k_z$,
then $\gamma$ comes to 1.03~mJ/mol-K$^2$.
\\

\begin{figure}[pt]
\includegraphics[width=3in]{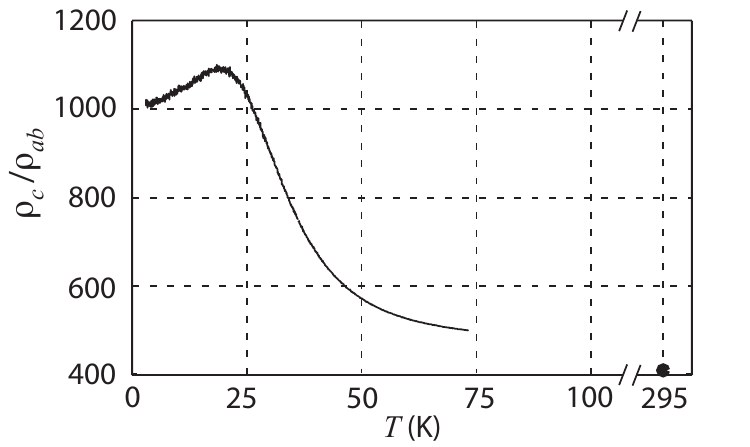}
\caption{\label{anisotropy}Resistive anisotropy $\rho_{c}/\rho_{ab}$ against temperature. The error from
sample geometrical factors is 10\%.}
\end{figure}

{\it \noindent Resistive anisotropy.} In the relaxation time approximation the expected anisotropy of a
corrugated cylindrical FS is given by:
\[
\rho_c/\rho_{ab} = \left[ d^2/2 \sum_{\mu, \nu >0} k_{\mu \nu}^2 (1+\delta_{\mu 0}) \right]^{-1} ,
\]
where $\delta_{\mu 0}$ is the Kronecker delta~\cite{Bergemann03} and $d$ the interlayer spacing. Using the
$k_{\mu\nu}$ from Table~1 of the main text yields $\rho_c/\rho_{ab}=241(13)$. The observed resistive
anisotropy is larger (\Fig{anisotropy}), especially below 25~K, where it rises to more than 1000. The
suppression of Umklapp processes described in the main text may play a role in enhancing the anisotropy, and
the larger anisotropy at low temperatures suggests that the defect scattering cross sections are themselves
anisotropic. Overall, however, the scale of the resistive anisotropy of PdCoO$_2$ should be the subject of
further investigation.
\\

{\it \noindent Tight-binding model.} In the text of the paper we describe how strong interlayer
next-nearest-neighbor ({\it nnn}) coupling, as would result from Co-mediated hopping, would give a large
$k_{31}$.  For confirmation, we performed a tight-binding calculation. The transfer integrals included are
in-plane {\it nn} and {\it nnn} coupling, $t_{nn} \equiv 1$ and $t_{nnn}$, and interlayer {\it nn} and {\it
nnn} coupling, $t_z$ and $t_{zz}$.  $t_{nnn}=-0.23$ and $E_F=0.22$ approximately (although not uniquely)
reproduce the half-filling and hexagonal shape of the Fermi surface. We calculate the FS for two sets of
parameters, and obtain the harmonic components:
\newline (1) $(t_{nn}, t_{nnn}, t_z, t_{zz})=(1, -0.23, 0.042, 0.011)$: 
\newline \hspace*{10mm} $k_{60}=0.039$, $k_{01}=0.0106$, $k_{31}=0.0011$~\AA$^{-1}$.
\newline (2) $(t_{nn}, t_{nnn}, t_z, t_{zz})=(1, -0.23, 0, 0.04)$: 
\newline \hspace*{10mm} $k_{60}=0.039$, $k_{01}=0.0082$, $k_{31}=0.0103$~\AA$^{-1}$.
\newline In the second set, where $t_{zz}$ dominates $t_z$, $k_{31}>k_{01}$, in disagreement with the
experimental data.
\begin{figure}[pb]
\includegraphics[width=3.25in]{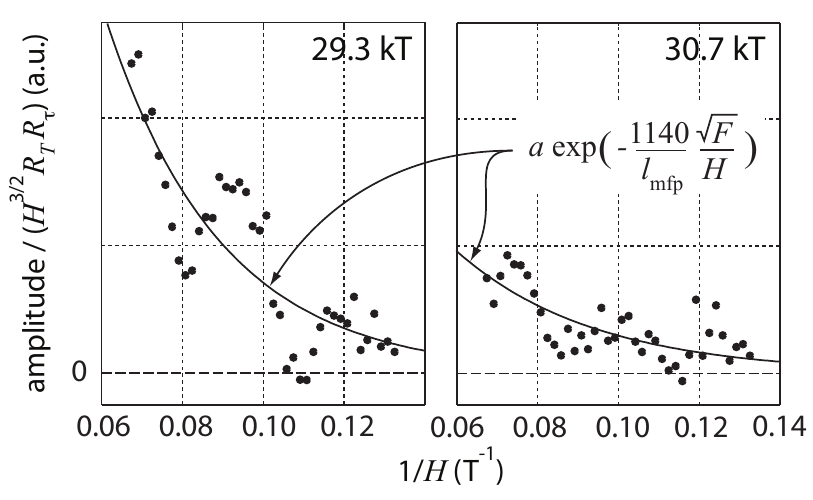}
\caption{\label{Dingle} Amplitudes of the 29.3 and 30.7~kT peaks against $1/H$, obtained by Fourier
transforming over windows of width 0.00167~T$^{-1}$. $R_T$ is the Lifshitz-Kosevich temperature attenuation
form, with $T=0.7$~K, and $R_\tau$ is the attenuation due to the measurement time constant. The solid lines
are fits to the standard Dingle attenuation; $F$ is the dHvA frequency in tesla, $l_{\mathrm{mfp}}$ the Dingle
mean free path in~\AA, and $H$ the field in tesla.}
\end{figure} 
\\

{\it \noindent Dingle analysis.} In attempting a Dingle analysis, we concluded that the dHvA amplitudes are
probably limited by phase smearing rather than impurity scattering: to observe a Dingle mean free path of
20~$\mu$m would require a field homogeneity better than one in $10^5$. Such high homogeneity would be
difficult to achieve even in the applied field, and PdCoO$_2$ can contain Co-based impurity phases that are
likely to increase the internal field inhomogeneity~\cite{Takatsu07, Tanaka96}. Our dHvA amplitudes, after the
effects of finite temperature and measurement time constant are removed, are shown in \Fig{Dingle}. Impurity
scattering should yield a straightforward exponential decay, and fits to the data yield Dingle mean free paths
of 5500 and 7000~\AA \ for the 29.3 and 30.7~kT frequencies, respectively. However the exponential fits do
not fit very well: the 29.3~kT amplitude shows pronounced minima at $\approx0.08$ and 0.11~T$^{-1}$, and the
30.7~kT amplitude appears to have minima at similar locations. This double-minimum form persists over a wide
range of $\theta$, and so is not due to a small FS warping: it is the reason that many of the frequencies in
\Fig{freqFit} are double peaks with a $\sim$60~T splitting. Nor is it due to magnetic interaction: it persists
to $T \sim 3$~K. Phase smearing is a possibility, with the minima being resonances where destructive
interference is maximum, however the origin of this possible phase smearing is unclear.


\begin{thebibliography}{99}
\bibitem{FSareaFootnote}Analytic expressions accurate to first order
in $k_{\mu\nu}$ are provided in~\cite{Bergemann03}, and more accurate expressions in~\cite{Grigoriev10}.
\bibitem{Kfoot}The $T^3$ term of the specific heat of potassium is 2.57~mJ/mol-K$^4$~\cite{Lien64}, yielding
$c=1290$~m/s.  $k_U=0.24$~\AA$^{-1}$, so $\hbar c k_U / k_B=24$~K, in good agreement with the activation
temperature obtained from fitting $\rho(T)$: 23~K~\cite{Gugan71}.
\bibitem{Dingle}A standard Dingle analysis of the dHvA oscillations yields a much lower mean free path of
0.6~$\mu$m~\cite{Supp}. However this result should be treated with caution, because as conductivity becomes
very high phase smearing due to internal field inhomogeneity becomes more problematic. Observing a Dingle mean
free path matching the tranport mean free path of 20~$\mu$m would require a field homogeneity of one in
$10^5$, probably difficult to achieve in a material with known Co-based, magnetic impurity
phases~\cite{Takatsu07}.
\end{thebibliography}
\end{document}